\documentclass[aps,prl,twocolumn,groupedaddress,showpacs]{revtex4}
\usepackage{graphicx}
\usepackage{amsmath}
\begin{document}

\title{The Fermi Surface Effect on Magnetic Interlayer Coupling}
\author{Erik Holmstr\"om}
\email[]{erik.holmstrom@fysik.uu.se}
\affiliation{Condensed Matter Theory Group, Physics Department,
Uppsala University, S-75121 Uppsala, Sweden}

\author{Anders Bergman}
\affiliation{Condensed Matter Theory Group, Physics Department,
Uppsala University, S-75121 Uppsala, Sweden}

\author{Lars Nordstr\"om}
\affiliation{Condensed Matter Theory Group, Physics Department,
Uppsala University, S-75121 Uppsala, Sweden}

\author{S.B. Dugdale}
\affiliation{H.H. Wills Physics Laboratory, University of Bristol,
Tyndall Ave, BS8 1TL Bristol,UK}

\author{I. Abrikosov}
\affiliation{Condensed Matter Theory Group, Physics Department,
Uppsala University, S-75121 Uppsala, Sweden}

\author{B.L. Gy\"{o}rffy}
\affiliation{H.H. Wills Physics Laboratory, University of Bristol,
Tyndall Ave, BS8 1TL Bristol,UK}

\date{\today}

\begin{abstract}
The oscillating magnetic interlayer coupling of Fe over spacer layers consisting 
of Cu$_{x}$Pd$_{1-x}$ alloys is
investigated by first principles density functional theory. The amplitude, period and phase of 
the coupling, as well as the 
disorder-induced decay, are analyzed in detail and the consistency to the 
Ruderman-Kittel-Kasuya-Yoshida (RKKY) theory is discussed. For the first time an effect
of the Fermi surface nesting strength on the amplitude is established from first principles 
calculations. An unexpected variation of the phase and disorder-induced decay is obtained 
and the results are discussed in terms of asymptotics.
\end{abstract}


\maketitle

\section{Introduction}

An interesting feature of random substitutional metallic alloys is their
rapidly but smoothly changing Fermi surfaces as the electron per atom 
ratio, $e/a$, varies with concentration. 
Such Fermi surface evolution can give rise to dramatic physical phenomena like 
spin- and charge-density waves or compositional ordering \cite{Ohshima73}, to 
mention but a few. 

A well-studied effect which is directly governed by the 
Fermi surface is the magnetic interlayer coupling (MIC) between two 
magnetic surfaces across a paramagnetic spacer as the spacer thickness 
is varied \cite{Bruno95,Stiles93,Stiles99,Niklasson96}. 
The theory of the MIC is well developed in the cases
of both pure metal and random substitutional metallic alloy spacers.
The common models are the so called Ruderman,
Kittel, Kasuya, Yoshida (RKKY)  model \cite{Bruno91,Bruno92}  and 
the quantum well (QW) model \cite{Stiles93}.  
The theories predict that among several things, the Fermi surface 
will play an important role in changing 
the period, amplitude, phase and decay of the MIC when the spacer is alloyed.
The phase is affected by the type of extremal points on the spacer 
Fermi surface, which may change with concentration.
The period changes since the length of the Fermi 
surface caliper changes as confirmed by Okuno 
\cite{Okuno93} and Bobo \cite{Bobo93} for a Co/Cu$_{1-x}$Ni$_{x}$/Co system 
and investigated theoretically by Lathiotakis 
\cite{Lathiotakis98a,Lathiotakis98b,Lathiotakis2000}. 
Finally, the amplitude of the MIC oscillation is influenced by the change in 
nesting at the Fermi surface.

In some cases, the amplitude does not change very much when the 
spacer is alloyed \cite{Bobo93,Okuno93} but in other alloys 
and for some growth directions the effect is dramatic 
\cite{Bruno97,Kudrnovsky96a,Takanashi96}. 
In the cases where the amplitude is changed by alloying
the spacer, it always becomes smaller with increasing impurity concentration.
In the studied materials, the decrease in amplitude is not a nesting effect 
but a disorder-induced damping of the electronic states in the spacer.

One very interesting case where the Fermi surface nesting could 
affect the amplitude of the MIC in addition to the disorder broadening 
is the Cu$_{x}$Pd$_{1-x}$ alloy. 
This system exhibits Fermi surface driven 
compositional ordering where the nesting of the 
Fermi surface is responsible for the concentration-dependent peaks observed 
in x-ray diffuse scattering in the concentration range 
$0.5 \le x \le 0.6$ \cite{Ohshima73,Gyorffy83}.   
Recent experimental studies of the Fermi surface nesting
 show an exceptionally flat region in the [110] direction in a
fcc Cu$_{0.6}$Pd$_{0.4}$  random alloy sample \cite{Wilkinson01}.
From this observation it is reasonable to believe that the nesting might
manifest itself as an increase of the amplitude at this concentration. 
In this article we will calculate the MIC of the 
Fe/Cu$_{x}$Pd$_{1-x}$/Fe system as function of $x$.
We will investigate the variation of the period, amplitude, phase and
disorder-induced decay with concentration for spacer thicknesses up to 22 ML.
The nesting effect on the amplitudes will be analyzed in detail and the validity
of extracting asymptotic properties from this type of calculations is discussed.

\section{Theory}

\subsection{Definition}

In this paper, the following definition of the magnietic interlayer coupling (MIC)
was used:
\begin{equation}
J(N)=E^{tot}_{\uparrow \downarrow}(N)-E^{tot}_{\uparrow \uparrow}(N).
\end{equation}

Here $E^{tot}_{\uparrow \downarrow(\uparrow \uparrow)}$ is the 
total energy of the system with the total magnetic 
moment of the Fe layers on one side of the system antiparallel (parallel) to the 
Fe layers on the other side and $N$ the number of atomic monolayers in the 
spacer. 

In all calculations we used the 
Korringa, Kohn and Rostocker (KKR) \cite{Andersen94} method within
the frozen core and atomic-sphere
approximations (ASA) together with the local spin density
approximation as parameterized in ref. \cite{Pbe96}. To carry out the 
multilayer calculations, the interface Green's function technique 
developed by Skriver and Rosengaard \cite{Skriver91} was used. The 
bulk alloys as well as the layered alloys were treated within the 
coherent potential approximation (CPA) \cite{Soven67,Abrikosov93,Gyorffy72}.

An advantage of the Green's function technique is that it ensures a correct
description of the loss of translational symmetry perpendicular to
the interface without the use of an artificial slab or supercell
geometry. The multilayer systems consisted of self-consistently calculated bulk
potentials for fcc Fe as boundary conditions to the left and right of 
the multilayer region that consisted of the alloy spacer and some Fe layers
that were included in the self-consistent calculation. The spin 
alignment of the two 
sides was either parallel or antiparallel. The spacer material was a
disordered binary alloy of the form Cu$_{x}$Pd$_{1-x}$ for x=0.4-0.9. 
The calculations were converged up to an energy difference of 
0.1 $\mu$Ry between iterations. The k-point sampling convergence was 
checked, and we used 1024 k-points in the irreducible part of the 
two dimensional Brillouin zone (2dBZ).  
The bulk, as well as the multilayer calculations were 
calculated in an ideal fcc lattice with the lattice parameter linearly 
interpolated between Cu and Pd for each concentration. 
This means that the fcc Fe boundary conditions were re-calculated for 
every Cu concentration that was going to be used in the slab in order to 
adapt to the global volume change. The choice of fcc Fe in the structure 
is purely technical in order to optimize the speed of the calculations.
A more realistic system would be embedded, thin fcc Fe layers in the alloy but 
that choice would demand calculations that include more atomic layers 
and our investigation would become intractable. The MIC should, however, not be 
qualitatively affected by our choice of semi-infinite fcc Fe as boundary conditions 
since the properties of the MIC are mainly dictated by the spacer material.

All our calculations were performed scalar relativisticly and the detailed form 
of the Fermi surfaces may have changed if the calculations would have included spin-orbit
coupling. However, by comparing our bulk calculations of extremal Fermi surface vectors 
and shapes of the Fermi surfaces to the fully relativistic calculations and experiments 
in refs. \cite{Wilkinson01,Comment1}, we conclude that the error from the scalar 
relativistic approximation is small.

\section{Results}

\begin{figure}
    \includegraphics*[angle=0,width=0.45\textwidth]{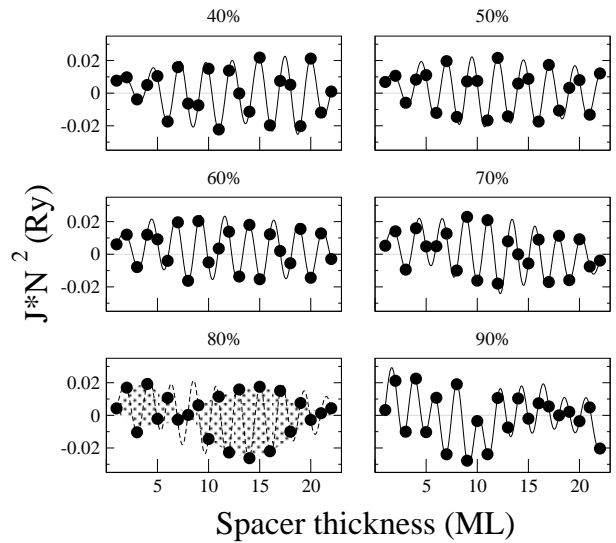}
    \caption{The MIC as function of spacer thickness for all the Cu concentrations considered 
                  (circles). The solid lines are the Fourier back transforms and they serve as 
                  a guide to the eye. The aliasing phenomena is visible for concentrations 
                  over 55\%. The occurrence of a second, longer period is visible in 
                  the 90\% case.   
    \label{fig:JNN}}
\end{figure}

\begin{figure}
\includegraphics*[angle=0,width=0.45\textwidth]{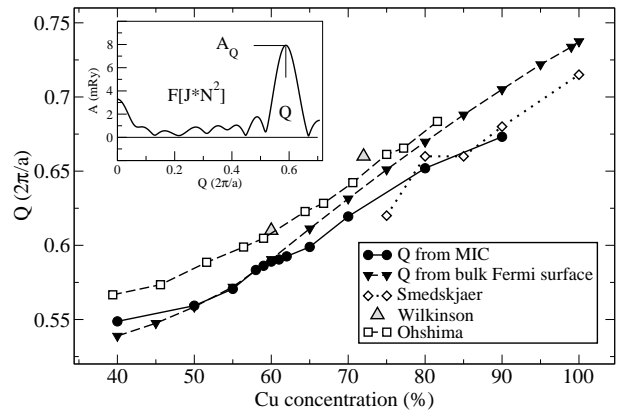}
\caption{The nesting vector as obtained by a direct 
         calculation of the bulk Fermi surface and as calculated from
         a Fourier transform of the MIC. The direction is (110). 
         Several experimental results
         are included for comparison: Smedskjaer \cite{Smedskjaer87},
          Wilkinson \cite{Wilkinson01} and Ohshima \cite{Ohshima73}. The inset 
         shows an example of a raw data Fourier transform of the MIC for 
         60\% Cu and how the values for the amplitude (A$_{Q}$) and nesting 
         vector (Q) are obtained.
    \label{fig:Q-conc}}
\end{figure}

\subsection{Magnetic moments}

The magnetic moments of the interface Fe layer is about 2.9$\mu_{B}$ for the 
40\% Cu systems and decreases linearly with concentration to 2.6$\mu_{B}$ 
for the 90\% Cu cases. This is mainly a volume effect since the global volume 
decreases linearly over the concentration range. 
For each Cu concentration, the change in moment of the interface Fe is very small
when the spacer thickness is varied.
In the spacer, the interface Pd atoms have a moment of 0.25$\mu_{B}$ for the 
40\% Cu system and 
0.18$\mu_{B}$ for the 90\% Cu case. The Cu atoms always have a very 
low moment, less than 0.05$\mu_{B}$. The layer-resolved magnetic 
moment in the spacer averaged over the Cu and Pd atoms always decays to zero within
4 ML from the Fe interface.

\subsection{Magnetic interlayer coupling}

In fig. \ref{fig:JNN} we have plotted the MIC for some of the Cu 
concentrations considered. In order to make the 
small oscillations visible, the amplitudes are multiplied by the square of the 
spacer thickness. 
We can see that the periods of the oscillations are between 2 and 3 ML and that 
no additional damping to the amplitudes is evident. There is also an obvious aliasing
effect in the 80\% case where the ``beat'' of the oscillation comes from the fact that 
the period is close to 2 ML and thus the frequency is close to the Nyquist frequency 
\cite{Nyquist28}. For the 80\% case, the beat is shown through shading,
but the effect is present for concentrations down to 55\% and we
believe that this phenomena is partly responsible for the uncertainty in the Fourier analysis
performed below. In the 90\% case there is also a new period that appears and this
can be seen from the ``wavy'' form of the MIC.
Further processing of the  
data is not possible without the aid of Fourier analysis and in the following, we will 
extract information from the Fourier spectra of the data in fig. \ref{fig:JNN}.

\subsection{Nesting vector}
      	
First we investigate the change in Fermi surface nesting vector as function of 
concentration.
In fig. \ref{fig:Q-conc} we have plotted the Fermi surface spanning vectors 
in the interval 0.4 $\le x \le$ 0.9 as obtained both directly
from the Fermi surface calculation and from the Fourier 
transform of the MIC as function of spacer thickness. The Fourier transforms always 
showed one single distinct peak and the q-vector for the peak could easily be 
obtained. A representative Fourier transform for $x=0.6$ is displayed in the inset. 
Although some of the layer thicknesses that were used in the Fourier transform 
clearly are not in the asymptotic region we still get a very good agreement with the 
nesting vectors from the bulk calculations.
We can see that the two theoretical curves agree within 5\% for all concentrations
which indicates that the Fermi surface is well defined in the multilayer system
despite the fact that the symmetry is broken in the direction of growth. 
The spanning vector increases from $\sim$0.55 to $\sim$0.67 within the considered 
concentration interval which translates into a period decrease of the MIC from 
$\sim$2.6 ML to $\sim$2.1 ML, respectively. For comparison, experimentally obtained
spanning vectors are also plotted, and the agreement is very good.

It is noteworthy that in our bulk calculation of Fe, we see a transition 
from a high-spin state to an intermediate-spin state where the magnetic moment changes from 
$\sim$2.5$\mu_{B}$  to $\sim$1.6$\mu_{B}$ when the 
lattice parameter is decreased below 3.61 \AA. This implies that we have
to limit our investigation to a concentration interval below $x=0.9$ in order
to avoid the effect of this transition in Fe on the MIC.

\begin{figure}
\includegraphics*[angle=0,width=0.45\textwidth]{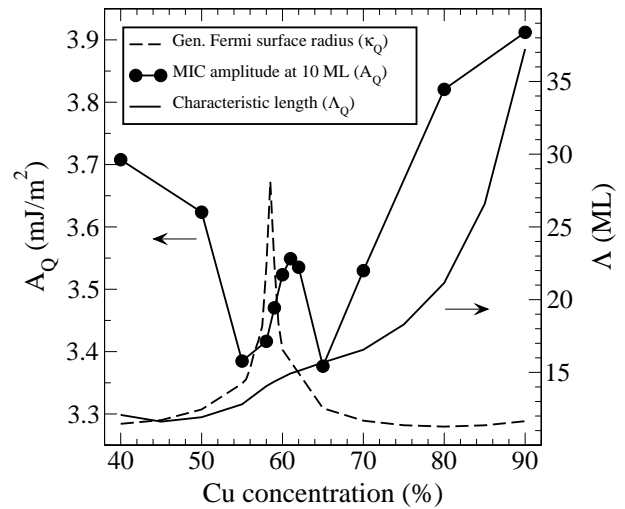}
\caption{The amplitude of the largest Fourier transform peak of the MIC for each 
         concentration together with the characteristic lengths and the Fermi surface 
         curvature as calculated 
         from eq. (\ref{eq:kappa}) (arb. units). 
    \label{fig:A-conc}}
\end{figure}

\subsection{Amplitude}

In this section we discuss the results obtained for the amplitudes associated 
with the spanning vectors shown in the previous section. 
In fig. \ref{fig:A-conc} we show the amplitude as calculated from the 
maximum of the Fourier transform for each Cu concentration. Also shown in the figure is
the generalized Fermi surface radius and the characteristic length 
displayed (which are two bulk properties that are defined by the radius associated with the 
curvature of the Fermi surface
and the inverse of the disorder-induced damping). For more details, see the
discussion below. 
The well-pronounced peak in the amplitude at $x=0.61$ is located 
in a global
minimum of the coupling which is in agreement with an increased characteristic length 
for concentrations to the right of the peak. 
The amplitude should, however, have a global minimum at 50\% Cu if only the characteristic 
length is considered, but in this alloy the $d$-band of Pd is
intersecting the Fermi energy for concentrations below 50\% and we believe this
changes the magnetic properties of the spacer, thus having a 
considerable effect on the MIC \cite{Holmstrom2003}.

We have also noticed that the Fourier transforms change
somewhat if we exclude the calculated points for the smallest spacers. However, 
the peak in the amplitude does not change position by more than $\pm$1 
on the concentration axis.
This means that despite not being in the asymptotic region, the effect of Fermi surface
nesting is clearly visible.

There are no experimental data concerning the MIC for systems with a CuPd 
alloy as a spacer and the only investigation of the Fermi surface 
nesting is that of Wilkinson \cite{Wilkinson01} by positron
annihilation. From that work, a crude estimate of the 
change in nesting may be obtained by comparing the number of measured nesting 
vectors from the total histograms of the Cu$_{60}$Pd$_{40}$ and 
Cu$_{72}$Pd$_{28}$ measured Fermi surfaces, which is a change of about 75\%. The 
change in MIC amplitude in our calculation between adjacent concentrations 
with a large difference in amplitude is about 6\%. 
To make a direct experiment on the MIC in this system would probably be a 
delicate task but the calculated change in amplitude is in principal not 
beyond experimental detection. We have 
calculated the MIC by assuming semi-infinite Fe layers in the fcc structure 
for computational reasons but the effect should also be seen in a system with embedded 
Fe or Co layers that could adapt to the fcc structure of the CuPd alloy.
Whether the fabrication of such multilayers is possible is, 
at least to our knowledge, an open question.

\subsection{Phase}
As explained in the general theory for a pure metal spacer in ref. \cite{Bruno92}, there should
be a phase-shift of the MIC associated with a change in the spacer Fermi surface curvature.

\begin{figure}
    \includegraphics*[angle=0,width=0.45\textwidth]{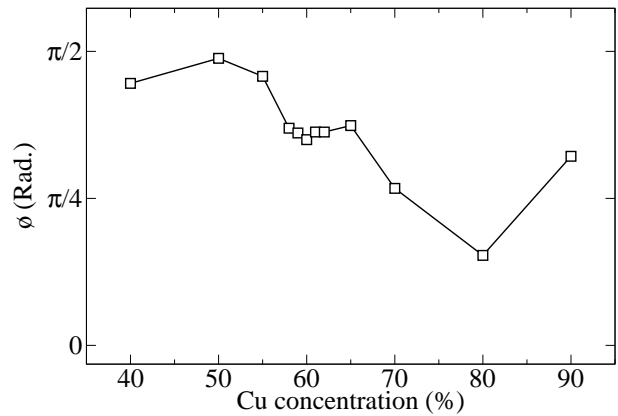}
    \caption[]{The phase shift of the oscillation 
                    as calculated by $\phi=\arctan(Im(F)/Re(F))$ 
		    and shifted back to 
                    the interval $0 \le \phi \le \pi$.
                    \label{fig:P-conc}}
\end{figure}

In our case, the neighbourhood around the nesting vector Q changes from a 
minimum to a saddle point when the Cu concentration is changed from 40\% to 90\%.
In order to investigate such a phase-shift in our calculation, we have calculated the phase 
($\phi$) of the oscillation ($J(N)$) from the Fourier transform ($F[J*N^{2}]$) by 
$\phi=\arctan(Im(F)/Re(F))$ and the result is shown in fig. \ref{fig:P-conc}.

It is clearly evident that our phase changes continuously with concentration and does not 
show any abrupt changes as might be expected. An explanation may be that the phase is 
more sensitive to the change in band matching at the interfaces as the concentration is varied
than to the change in Fermi surface curvature. Experimental studies of the phase 
as function of impurity concentration in the magnetic layer argue that the large observed 
phase change is due to the altered band matching at the interfaces \cite{Ebels98}.
In the alloy spacer, the Fermi surface is 
also not as well defined as in a pure metal (c.f. fig. \ref{fig:BSF})  and 
the diffuseness may then be responsible for a smearing of the phase shift over a 
much broader concentration range. There may also be other effects that influence the phase
such as the electronic topological transitions (ETT) of the Fermi surface at 
50\% and 63\% Cu \cite{EBruno2001}  and the aliasing effect due to the discrete monolayer 
sampling of the MIC.

\section{Discussion}

\subsection{Nesting from bulk Fermi surface}

In order to investigate the nesting from the bulk Fermi surface
we have adopted the spanning vector counting method suggested 
by Wilkinson \cite{Wilkinson01}. In order to do so, we calculated the 
spectral function in the full Brillouin zone according to eq. (\ref{eq:BSF})
on a  grid of (64x64x64) k-points. The spectral function was then 
further interpolated to a 120x120x120 mesh. This
mesh was used to construct an isosurface for a given intensity-cut of the
spectral function. This value was chosen as the highest value possible
that still resulted in a continuous surface. Since the isosurface was
constructed from the spectral function, it consisted of two separate
sheets, but the distance between those sheets was minimized due to the
choice of the intensity-cut and tests were made to ensure no double peaks
when the nesting check was calculated.
After this procedure the number of points on the Fermi surface
was about 70000. 

We then created a histogram of vectors connecting
two points on this surface along a given direction, in our case [110].
This histogram then showed a peak for
the vector length that is most frequently represented on our Fermi surface. 

The discrete representation of  the Fermi surface and the rounding error 
when calculating the length of the vectors with this method resulted 
in rough histograms that were smoothed by convolution with a gaussian function.
The intensity of the histograms was also normalized 
with the number of points on the Fermi surface in order to compare intensities
of different concentrations.

\begin{figure}
    \includegraphics*[angle=0,width=0.45\textwidth]{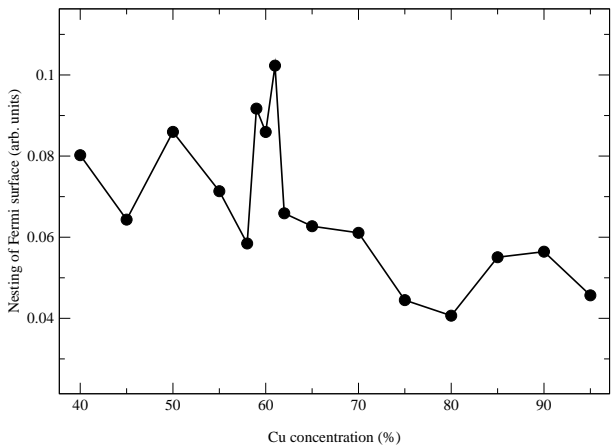}
    \caption[]{The normalized number of vectors that take part in the 
                    nesting in the [110] direction.
                    \label{fig:Nesting}}
\end{figure}

In fig. \ref{fig:Nesting} we have plotted the intensity maximum of the histogram
divided by the total number of points on the Fermi surface. It is clear that there 
is a peak for concentrations around 61\%. Compared to the maximum of
the generalized curvature there is a difference of about 3 units on the concentration 
axis and the result agrees perfectly with the maximum amplitude of the MIC.

\subsection{Model}

The strength of a total energy calculation is that 
the MIC is obtained directly from the independently calculated energies for each 
magnetic configuration by using 
the definition. To gain physical insight, however, 
we need to consider a model for the MIC. A good choice in this case is to look 
at the RKKY model for a simple quantum well potential but for a 
general Fermi surface. 
Such a case is described in detail in ref.\cite{Stiles93} where the final result
in the asymptotic limit  ($\infty$) is

\begin{equation}\label{eq:J(N)}
J_{\infty}(N)=\sum_{i} -\left [\frac{\hbar}{2\pi^{2}}\kappa_{i}\nu_{i} 
              |\Delta R|^{2} \right ] 
              \frac{\cos(Q_{i}N + \phi_{i})}{N^{2}}.
\end{equation}

\noindent
Here, the sum is over the stationary points of the Fermi surface and $Q_{i}$ are the 
vectors on the Fermi surface that connects the extremal points. 
The generalized Fermi surface radii $\kappa_{i}$ are defined as

\begin{equation}\label{eq:kappa}
\kappa_{i} = \left \lbrack \sqrt{\frac{\partial^{2}Q_{i}({k_{\parallel}})}{\partial k_{x}^{2}}
 \frac{\partial^{2} Q_{i}({k_{\parallel}})}{\partial k_{y}^{2}} - \left ( 
 \frac{\partial^{2} Q_{i}({k_{\parallel}})}{\partial k_{x}\partial k_{y}} \right )^{2}} 
\right \rbrack^{-1}
\end{equation}

\noindent
and  $\nu_{i}$ are the reduced Fermi velocities. 
$\Delta R = R^{\uparrow}_{\uparrow}-R^{\uparrow}_{\downarrow}$ 
is the difference in reflection amplitude between a
spin-up  and a spin-down electron in the well 
reflecting on a spin up barrier and by symmetry 
$R^{\uparrow}_{\uparrow}=R^{\downarrow}_{\downarrow}$  and 
$R^{\uparrow}_{\downarrow}=R^{\downarrow}_{\uparrow}$.

\begin{figure}
\includegraphics*[angle=0,width=0.45\textwidth]{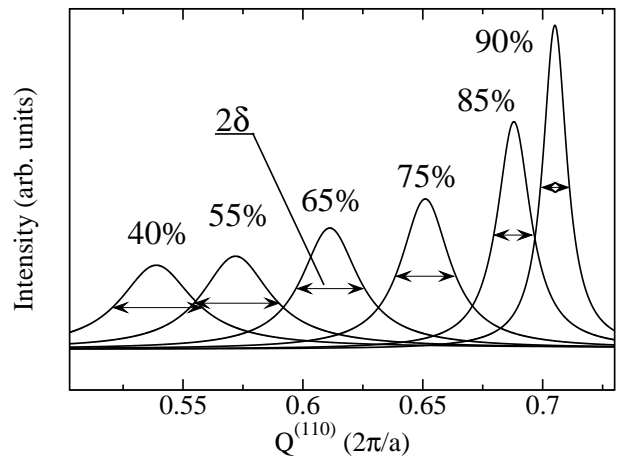}
\caption{The Bloch spectral function along the line $\Gamma-\Delta-X$
         for some of the calculated concentrations.
         The widths for each concentration  (2$\delta$) are indicated 
         by the horizontal arrows.
    \label{fig:BSF}}
\end{figure}

One could expect that the theory breaks 
down in case of an alloy spacer but as showed in references 
\cite{Bruno97,Lathiotakis98a} the effect of an alloy spacer is an additional 
exponential damping factor to the formula for the MIC so that

\begin{equation}
J^{alloy}_{\infty}(N)=J_{\infty}(N)e^{-\frac{N}{\Lambda}}.
\label{eq:J+damp}
\end{equation}

The characteristic length $\Lambda$ is given by

\begin{equation}
\frac{1}{\Lambda}=\frac{1}{\lambda^{+}}-\frac{1}{\lambda^{-}}
\end{equation}

\noindent
where $\lambda^{+(-)}$ are the mean-free-paths in the direction of 
growth at the two edges of the Fermi surface.
In our case we have a symmetric, single sheet Fermi surface so that the 
condition $\lambda^{+}=-\lambda^{-}$ is fulfilled and
the characteristic length can be calculated as $\Lambda=(\lambda^{+})/2$.  
The mean-free-paths are calculated as $1/\lambda = \delta$ where
$\delta$ is the half-width of the Fermi surface as illustrated in fig. \ref{fig:BSF}.

For the calculation of the Fermi surface half-widths, the Bloch spectral 
function

\begin{equation}\label{eq:BSF}
A^{\sigma}({\bf k},E)=-\frac{1}{\pi}\mathrm {Im Tr} \ G^{\sigma}({\bf k},E)
\end{equation}

\noindent
from the Greens function of spin $\sigma$, wave vector ${\bf k}$ and energy $E$ 
was evaluated at the Fermi energy $E_{F}$. Since we are only interested in the 
behavior along the [110] direction, we need not to perform
the calculation in the full Brillouin zone (BZ) but restrict the values of 
${\bf k}$ to the $\Gamma-\Delta-X$ line.

\begin{figure}
\includegraphics*[angle=0,width=0.45\textwidth]{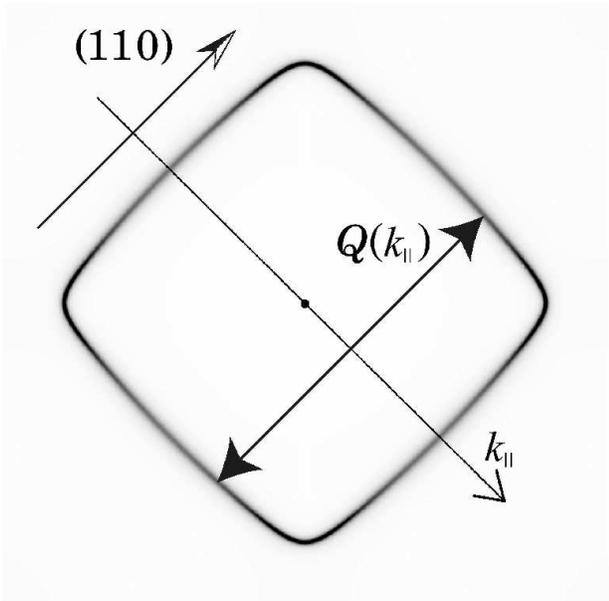}
\caption{The Fermi surface of Cu$_{0.7}$Pd$_{0.3}$ in the (001) plane. 
         The spanning vector $Q$ is displayed together with the direction of 
         growth ([110]) and the definition of $k_{\parallel}$.
    \label{fig:EF}}
\end{figure}

From now on we will also assume that there is only one
extremal spanning vector ($Q$) on the Fermi surface in the direction that we are investigating.
In fig. \ref{fig:EF}, a general spanning vector is displayed on a cut through the 
Cu$_{0.7}$Pd$_{0.3}$ Fermi surface. The generalized Fermi surface radius is 
calculated by using such spanning vectors in a centered difference approximation 
of eq. (\ref{eq:kappa}).

\subsection{Amplitude}

The amplitude of the MIC is a much more complicated quantity to calculate
compared to the period of the oscillation. From eq. (\ref{eq:J(N)}) we can see
that the amplitude, even in the ideal model case, depends on a number of 
factors. In this discussion we will assume that the Fermi velocity term ($\nu$)
is constant or at least very slowly varying over the concentration interval. An
estimation of the change in Fermi velocity may be performed by inspecting the 
bulk band structure of Cu and Pd for the [110] direction and taking the slope of
the band at the Fermi level. Our estimation gives a change in Fermi velocity 
of no more than 4\% over our concentration interval.

The reflection coefficients are much harder to estimate. They may, in general, 
vary irregularly with concentration and the quadratic contribution to 
the amplitude is strong. A quantitative investigation of the
variation of the reflection coefficients with concentration is a comprehensive
task and is beyond the scope of this article. Calculations of reflection coefficients 
were made in refs. \cite{Stiles96,Wildberger98,Riedel2001}. However, above $x=0.55$,
the d-band of Pd is already below the Fermi energy and the variation in band 
matching (at least at the Fermi level) is only from the s-band. The s-band is not 
changing very much between Cu and Pd so we assume that the band matching,
which gives the reflection coefficients, does not change very much in the interval.
We therefore assume that the reflection coefficients will not affect the trend of 
the amplitudes more than in a monotonic way.

The remaining quantities that affect the coupling are the 
generalized Fermi surface radius and characteristic length.
They were calculated from the Fermi 
surfaces of the corresponding CuPd bulk alloys and since we are not 
completely in the asymptotic region, the comparison 
to the amplitudes from the full multilayer calculations 
is not strictly justified. However, the concentrations where $\kappa$ diverges and 
the amplitude has a maximum are very close and we argue that the small discrepancy
is partly due to this pre-asymptotic effect. 

Close to $x=0.58$, the Fermi surface is perfectly flat in one $k_{\parallel}$
direction and does thus show nesting along a line. The result is that 
eq. (\ref{eq:kappa}) breaks down and diverges. At the pole $\kappa$
changes sign which reflects the change from a minimum to a saddle point 
around the nesting caliper. In fig. \ref{fig:A-conc}, therefore, the absolute 
value of $\kappa$ is displayed (in arbitrary units).

\subsection{Decay}
The characteristic length that was calculated within this model is displayed in 
fig. \ref{fig:A-conc}. In order to check for the decay that is associated with 
the characteristic length, we have also performed a least squares fit of exponentially decaying 
functions to the MIC. From 
that analysis, we have concluded that there is no decay in the calculated MIC on the 
order of the estimated decay in fig. \ref{fig:A-conc}. This indicates that the decay 
is not easily observed in the studied spacer thickness range.

The lack of damping is also evident in two cases in ref. 
\cite{Kudrnovsky96a}, fig. 3 where the MIC was calculated for Cu$_{0.5}$Au$_{0.5}$,
Cu$_{0.75}$Ni$_{0.25}$ and Cu$_{0.5}$Zn$_{0.5}$. 
We have calculated the characteristic lengths for these three alloys in the same way as for our
CuPd case to be  23-,87- and 23 ML respectively.
In ref. \cite{Kudrnovsky96a}, there is only clear 
exponential damping for the CuZn case although the damping is the same for 
the CuAu alloy. It is then very 
interesting to examine ref. \cite{Bruno97} where the same authors present an 
extended calculation of the CuAu system (fig. 4e) where they double the number of calculated 
spacer layers from 45 to 90ML, the damping then appearing for thicknesses over 45 ML.
An estimate of the exponential damping from the figures presented gives 
the characteristic length for the CuAu case to be $\sim$76ML whereas the same 
property for CuZn becomes $\sim$25ML which agrees with our calculated characteristic 
length from the Fermi surface.
Thus, the damping term may not 
be as simple as previously thought and may contain some unknown, element specific, 
prefactor which would
explain the appearance of the damping in the CuAu case. It may also be that, 
for some cases, the damping cannot be calculated from 
a single point on the Fermi surface by using the Bloch spectral function within the CPA.

Since we calculate the characteristic lengths from the point where the nesting vector 
touches the Fermi surface we neglect contributions from all other vectors when the 
Fermi surface is flat. In our case this may be a large source of error since we are 
investigating a system with substantial nesting.
The lack of damping may also be of unphysical origin. It is well known that the MIC 
may diverge in an exponential way for large spacer thicknesses if the number of k-points
is too low in total energy calculations. 
It may then be the case that for a certain number of k-points, the exponential divergence is 
canceled by the damping so that the total result appears converged.
It is not known how all factors in 
eq. (\ref{eq:J(N)}) converge with k-points for total energy calculations and we 
speculate that the exponential damping term is very hard to converge. The disorder 
induced damping is also quite weak compared to the normal $1/N^{2}$ decay. 
As an example, we have estimated the ratio of the exponential damping to the 
$1/N^{2}$ decay in our CuPd system to be 77 times larger for a spacer thickness 
of 22 ML.

\section{Conclusions}

The amplitude maximum of the MIC and the maximum nesting strength 
show a remarkable agreement. We thus conclude that the MIC is affected
by the nesting in a way that is well described by the RKKY model. 
However, the agreement between the divergence of the generalized 
Fermi surface radius and the peak in MIC is not perfect and an analysis of the 
nesting of the bulk alloy Fermi surfaces show that the true nesting peak 
and the divergence in Fermi surface radius do not occur at exactly the 
same concentration. 

The expected phase-shift that is associated with the divergence of the generalized 
Fermi surface curvature is not seen in our calculations but the phase changes
continously over the concentration range. We expect that the phase-shift 
should be seen if the calculation was extended further into the asymptotic region.
We also do not see the anticipated disorder-induced decay of the amplitude 
and the comparison to calculations by Bruno and Kudrnovsky 
\cite{Bruno97,Kudrnovsky96a}
indicates that this decay may be visible only for very large systems (N$>45$).

The increase in the calculated amplitude is about 10\% and would in principle be
measurable in an experiment if the interface quality is good enough. 
 
\section{Summary}
We have performed full {\it ab initio}, total energy calculations of the MIC in 
Fe/Cu$_{x}$Pd$_{1-x}$/Fe random alloy systems for 0.4 $\le x \le 0.9$ and
spacer thicknesses of 1-22 ML. 
At the concentration $x\sim0.6$ we see a large effect on the 
amplitude from Fermi surface nesting. We have also 
investigated the period, phase and disorder-induced decay of the MIC. 
The small difference in 
predicted amplitude maximum from bulk Fermi surface calculations is argued to 
originate mainly from pre-asymptotic effects. 
The results give important information
on the applicability of asymptotic models for the MIC.

\begin{acknowledgments}
This work was supported from the Swedish Research Council (VR), 
the Swedish Foundation for
Strategic Research (SSF) and the European Network for Computational 
Magnetoelectronics. SBD is supported by the Royal Society (UK).
Discussions with Nektarios Lathiotakis and Josef Kudrnovsky 
are gratefully acknowledged. Special thanks also goes to B. Liz$\rm\acute{a}$rraga 
for strong support.
\end{acknowledgments}


\end{document}